\documentclass[aps,prl,nofootinbib,twocolumn,superscriptaddress,floatfix,notitlepage]{revtex4-1}
\pdfoutput=1

\usepackage{graphicx}
\usepackage{amsmath,amsfonts,amssymb}
\usepackage{color}
\usepackage[breaklinks,colorlinks,urlcolor=blue,citecolor=blue,linkcolor=magenta]{hyperref}
\usepackage{verbatim}
\usepackage{enumitem}
\usepackage{hyperref}
\usepackage{natbib}
\usepackage{aas_macros}

\usepackage[table]{xcolor}
\usepackage{array}
\newcolumntype{L}[1]{>{\raggedright\let\newline\\\arraybackslash\hspace{0pt}}m{#1}}
\newcolumntype{C}[1]{>{\centering\let\newline\\\arraybackslash\hspace{0pt}}m{#1}}
\newcolumntype{R}[1]{>{\raggedleft\let\newline\\\arraybackslash\hspace{0pt}}m{#1}}
\newcommand{\be}{\begin{equation}}
\newcommand{\ee}{\end{equation}}
\newcommand{\bea}{\begin{equation}\begin{aligned}} 
\newcommand{\eea}{\end{aligned}\end{equation}}
\newcommand{\td}{{\rm d}}

\newcommand{\Msun}{M_\odot}
\newcommand{\vect}[1]{\boldsymbol{#1}}

\def\lsim{\mathrel{\raise.3ex\hbox{$<$\kern-.75em\lower1ex\hbox{$\sim$}}}}
\def\gsim{\mathrel{\raise.3ex\hbox{$>$\kern-.75em\lower1ex\hbox{$\sim$}}}}

\begin{document}

\title{Consistency of JWST Black Hole Observations with \\
\vspace{1mm}
NANOGrav Gravitational Wave Measurements}

\author{John Ellis}
\email{john.ellis@cern.ch}
\affiliation{Physics Department, King’s College London, Strand, London, WC2R 2LS, United Kingdom}
\affiliation{Theoretical Physics Department, CERN, CH 1211 Geneva, Switzerland}

\author{Malcolm Fairbairn}
\email{malcolm.fairbairn@kcl.ac.uk}
\affiliation{Physics Department, King’s College London, Strand, London, WC2R 2LS, United Kingdom}

\author{Gert H\"utsi}
\email{gert.hutsi@kbfi.ee}
\affiliation{Keemilise ja Bioloogilise F\"u\"usika Instituut, R\"avala pst. 10, 10143 Tallinn, Estonia}

\author{Juan Urrutia}
\email{juan.urrutia@kbfi.ee}
\affiliation{Keemilise ja Bioloogilise F\"u\"usika Instituut, R\"avala pst. 10, 10143 Tallinn, Estonia}
\affiliation{Department of Cybernetics, Tallinn University of Technology, Akadeemia tee 21, 12618 Tallinn, Estonia}

\author{Ville Vaskonen}
\email{ville.vaskonen@pd.infn.it}
\affiliation{Keemilise ja Bioloogilise F\"u\"usika Instituut, R\"avala pst. 10, 10143 Tallinn, Estonia}
\affiliation{Dipartimento di Fisica e Astronomia, Universit\`a degli Studi di Padova, Via Marzolo 8, 35131 Padova, Italy}
\affiliation{Istituto Nazionale di Fisica Nucleare, Sezione di Padova, Via Marzolo 8, 35131 Padova, Italy}

\author{Hardi Veerm\"ae}
\email{hardi.veermae@cern.ch}
\affiliation{Keemilise ja Bioloogilise F\"u\"usika Instituut, R\"avala pst. 10, 10143 Tallinn, Estonia}

\begin{abstract}
JWST observations have opened a new chapter in supermassive black hole (SMBH) studies, stimulating discussion of two puzzles: the abundance of SMBHs in the early Universe and the fraction of dual AGNs. In this paper, we argue that the answers to these puzzles may be linked to an interpretation of the data on the nHz gravitational wave (GW) discovered by NANOGrav and other pulsar timing arrays (PTAs) in terms of SMBH binaries losing energy by interactions with their environments as well as by GW emission. According to this interpretation, the SMBHs in low-$z$ AGNs are the tip of the iceberg of a local SMBH population populating mainly inactive galaxies. This interpretation is consistent with the ``little red dots" observed with JWST being AGNs, and would favour the observability of GW signals from BH binaries in LISA and deciHz GW detectors.
\\~~\\
KCL-PH-TH/2024-16, CERN-TH-2024-038, AION-REPORT/2024-03
\end{abstract}

\maketitle

\vspace{5pt}\noindent{\bf Introduction --}  There has been some surprise at the number of supermassive black holes (SMBHs) discovered with the JWST at high redshifts $z$: see~\cite{2023A&A...677A.145U,2023ApJ...953L..29L,2023ApJ...959...39H, Bogdan:2023ilu, Ding_2023, Maiolino:2023bpi, 2023arXiv230904614Y}, the red data points shown in Fig.~\ref{fig:massrelations} and the scatter plot in the Appendix. Prior expectations for the number of SMBHs in the early Universe were largely based on empirical relationships between the SMBH mass-halo mass ratio and the SMBH mass-stellar mass ratio, such as that found by Reines and Volonteri (RV) in the local (low-$z$) population of active galactic nuclei (AGNs)~\cite{2015ApJ...813...82R}: see the green points in Fig.~\ref{fig:massrelations}. In a statistical analysis, Pacucci et al.~\cite{Pacucci:2023oci} concluded that there was a $> 3\sigma$ discrepancy between the SMBH mass/stellar mass ratios, $M_{\rm BH}/M_*$, of a subset of the high-$z$ JWST data and this fit to local AGN data. We note, however, that RV also analysed a sample of local inactive galaxies (IGs), finding higher values of $M_{\rm BH}/M_*$~\cite{2015ApJ...813...82R}: see the blue points in Fig.~\ref{fig:massrelations}. Moreover, larger values of $M_{\rm BH}/M_*$ had also been found in a compilation of high-$z$ SMBHs observed previously by JWST in~\cite{2021ApJ...914...36I}: see the yellow data in Fig.~\ref{fig:massrelations}.

Another surprise in JWST data was the discovery of several close pairs in a sample of AGNs~\cite{2023arXiv231003067P}, whereas cosmological simulations generally predicted a fraction of such dual AGNs that was at least an order of magnitude smaller. This observation was used in~\cite{Padmanabhan:2024nvv} to study SMBH binary evolution. JWST observations have also revealed a population of ``little red dots" (LRDs) in the redshift range $z \sim 2 - 11$ that are interpreted as previously-hidden AGNs tracing BH growth in the early Universe~\cite{2023Natur.616..266L,2023ApJ...956...61A,2024ApJ...963..128B,2023arXiv230607320L, Matthee:2023utn,2024arXiv240403576K}.

In parallel with the emergence of JWST data, pulsar timing arrays (PTAs) have reported the existence of a stochastic gravitational wave background (SGWB) in the nHz range~\cite{NANOGrav:2023gor, EPTA:2023xxk, Reardon:2023gzh, Xu:2023wog}, whose most plausible astrophysical interpretation is gravitational wave (GW) emission by SMBH binaries. Analyzing their data under this assumption and guided by astrophysical priors, the NANOGrav Collaboration found~\cite{NANOGrav:2023hfp} that their data were best fit by a $M_{\rm BH}/M_*$ ratio that was significantly higher than that found in the RV analysis of local AGNs~\cite{2015ApJ...813...82R}. In parallel, we found independently in~\cite{Ellis:2023dgf}, using the Extended Press-Schechter formalism~\cite{Press:1973iz, Bond:1990iw} to estimate the halo merger rate and hence the evolution of the halo mass function, and assuming a fixed probability, $p_{\rm BH}$, for a halo merger to yield a BH merger, that a good fit to the NANOGrav data was possible with the $M_{\rm BH}/M_*$ ratio found by RV in their analysis of local IGs~\cite{2015ApJ...813...82R}. It was also found in ~\cite{Ellis:2023dgf} that using the RV fit to local AGNs degraded significantly the fit to the NANOGrav data: $\Delta \chi^2 = 43$ compared to the fit based on local IGs: see Fig.~9 of~\cite{Ellis:2023dgf} and the discussion in the accompanying text, as well as the analysis below.

It has been suggested recently that the $M_{\rm BH}/M_{\rm *}$ relationship may evolve with $z$~\cite{Pacucci:2024ijt}, or that selection biases and measurement uncertainties~\cite{2024arXiv240300074L} may account for much of the tension between the JWST high-$z$ SMBH data and the data on local AGNs considered in~\cite{2015ApJ...813...82R}.
However, we take at face value the {\it prima facie} evidence for consistency between the information on SMBHs provided by the high-$z$ SMBH data and NANOGrav measurements. The puzzle of the masses of JWST black holes may be linked to the somewhat unexpected strength of the GW signal reported by NANOGrav and other PTAs~\cite{InternationalPulsarTimingArray:2023mzf}, and can be reframed as a puzzle why both data sets correspond better to the data on local IGs than to local AGNs.

In this paper we quantify in more detail the consistency between the NANOGrav nHz GW data and the JWST and previous data on high-$z$ SMBHs, discussing both the SMBH mass/stellar mass ratio and the fraction of dual AGNs in the JWST data. We confirm the tension between the SMBH mass-stellar mass ratio found in the population of  low-$z$ AGNs and a subset of the JWST data found in~\cite{Pacucci:2023oci}. However, we also show that the JWST data are much more compatible with the low-$z$ IG data, and that the JWST data are also compatible at an even higher level with the compilation of earlier SMBH data in~\cite{2021ApJ...914...36I}. Moreover, we show that the high-$z$ SMBH data are also consistent with our analysis of the NANOGrav data on the nHz GW background~\cite{Ellis:2023dgf}. Our interpretation of the NANOGrav and JWST data would favour the observability of GW signals from BH binaries in LISA and deciHz GW detectors, as discussed previously in~\cite{Ellis:2023dgf}. It can also accommodate the unexpectedly high fraction of dual AGNs observed with JWST~\cite{2023arXiv231003067P}. Assuming that the hardening timescale due to halo-halo interactions is similar to that found empirically for environmental effects in our fit to NANOGrav data~\cite{Ellis:2023dgf}, we find consistency at the 95\% confidence level, suggesting linkage also to the dual AGN puzzle-- which also favours the observability of GWs from massive BH mergers. In addition, our interpretation of the JWST and NANOGrav data is consistent with the population of ``little red dots" reported recently~\cite{Matthee:2023utn,2024arXiv240403576K}, assuming that they are AGNs triggered during mergers as discussed in more detail below.

\begin{figure}
    \centering
    \includegraphics[width=\columnwidth]{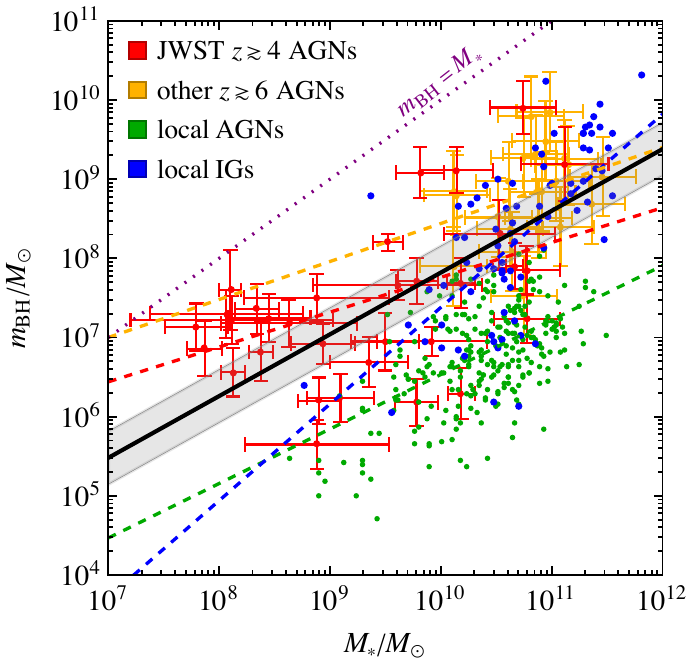}
    \caption{Measurements and fits of the black hole mass--stellar mass relation. The data points with errors are from JWST measurements (red, see the Appendix) and other high-$z$ measurements (yellow)~\cite{2021ApJ...914...36I}. The local AGN measurements (green) and the local IG measurements (blue), both from~\cite{2015ApJ...813...82R}, are shown without errors, but the errors are taken into account in the analysis. The dashed lines show the best fits to each of the data sets separately and the black solid line shows the best fit combining the high-$z$ data and the local IGs data. The grey band shows the scatter in the latter fit.}
    \label{fig:massrelations}
\end{figure}

\vspace{5pt}\noindent{\bf Analysis of the BH mass-stellar mass relation --}  We consider the $33$  high-$z$ AGNs observed by JWST compiled in the Appendix. These are displayed in red in Fig.~\ref{fig:massrelations} together with their measurement uncertainties. In addition, we also consider a sample of 21 high-$z$ AGNs observed previously, extracted from the observations compiled in~\cite{2021ApJ...914...36I}. These are displayed in yellow in Fig.~\ref{fig:massrelations}. The errors in the BH masses were reported in~\cite{2021ApJ...914...36I}, but not the uncertainties in the stellar masses: we approximate them by $0.3$\,dex. The blue and green points in Fig.~\ref{fig:massrelations} show, respectively, the local IG and AGN measurements compiled in~\cite{2015ApJ...813...82R}. For clarity, we do not show the error bars for these points, but we take the uncertainties reported in~\cite{2015ApJ...813...82R} into account in the following analysis.

The new JWST SMBH observations have been interpreted as inconsistent with the local mass relationship~\cite{Pacucci:2023oci}, prompting, e.g., the proposals that the SMBH-stellar mass relationship evolves with redshift~\cite{Pacucci:2024ijt}, or that selection biases and measurement uncertainties in the high-$z$ population may be largely responsible for the apparent tension with the local relationship~\cite{2024arXiv240300074L}. 
As mentioned earlier, rather than pursuing these hypotheses, we recall the tension between different local active and inactive SMBH populations~\cite{2015ApJ...813...82R} and emphasize the consistency of the JWST measurements with the local IG measurements as well as the other high-$z$ measurements.

As commonly done in the literature, we assume a power-law form for the relation between the SMBH mass and the stellar mass of the host galaxy, with a log-normal intrinsic scatter. The probability distribution of the SMBH mass $m$ for a given stellar mass $M_*$ is then parametrized by the magnitude $a$ and logarithmic slope $b$ of the mean and the width $\sigma$ of the distribution:
\be \label{eq:parameters}
    \frac{\td P(m|M_*,\vect{\theta})}{\td \log_{10} \!m} = \mathcal{N}\bigg(\!\log_{10} \!\frac{m}{M_\odot} \bigg| a + b \log_{10} \!\frac{M_*}{10^{11}M_\odot},\sigma\bigg) \, ,
\ee
where $\vect{\theta} = (a,b,\sigma)$ and $\mathcal{N}(x|\bar x,\sigma)$ denotes the probability density of a Gaussian distribution with mean $\bar x$ and variance $\sigma^2$.

We fit the parameters $a$, $b$ and $\sigma$ to the measured data. The likelihood function is
\bea
    \mathcal{L}(\vect{\theta}) &\propto \prod_j \int \!\td \log_{10} \!m \, \td \log_{10} \!M \, \frac{\td P(m|M,\vect{\theta})}{\td \log_{10} m}\\
    &\times \mathcal{N}(\log_{10} m|\log_{10} m_j) \mathcal{N}(\log_{10} M|\log_{10} M_j) \, ,
\eea
where the product is over the data points, $m_j$ and $M_j$ denote the mean measured BH and stellar masses, and we assume that the posteriors of the BH and stellar mass measurements are log-normal and uncorrelated. We use flat priors for $a$, $b$ and $\sigma$ with $b>0$.

As seen in Fig.~\ref{fig:massrelations}, the high-$z$ AGN measurements are largely consistent with each other and with the fit to the local IGs, though the latter prefers a slightly stronger tilt. The posteriors in $a$ and $b$ found in a global fit to these data, marginalized over $\sigma$, are shown in the upper panel of Fig.~\ref{fig:Ellipses}, and the posteriors of the fits to individual data sets are shown in the Appendix. The lower panel of Fig.~\ref{fig:Ellipses} displays uncertainty ellipses for the individual data sets as well as that for the global fit (shown in grey) that omits the local AGN data (shown in green). The global best fit is given by
\be \label{eq:globalfit}
    a = 8.6 \,, \quad b = 0.8 \,, \quad \sigma = 0.8\, \,,
\ee
and the 68\% CL parameter ranges are indicated by the vertical dashed lines in the upper panel of Fig.~\ref{fig:Ellipses}. The fit to the local AGN population prefers, instead, much lower $a$ values but has a slope $b$ similar to that in the global fit. We note that the linear relation ($b = 1$) is compatible with our global fit at the 68\% CL.

We quantify the overlaps between the posteriors for the fits to the different data sets $i$ and $j$ using the overlap coefficient
\be \label{eq:O}
    O_{i,j} = \int \td \vect{\theta} \min\left[\mathcal{L}_i(\vect{\theta}), \mathcal{L}_j(\vect{\theta}) \right] \,.
\ee
The overlap coefficients are tabulated in Table~\ref{tab:overlaps}. We see numerically that the fit to the JWST data has better overlaps with the fits to the other high-$z$ SMBHs and local IGs, as well as the global fit, than with the fit to local AGNs. The local AGN data also have very low overlaps with the other high-$z$ SMBHs, the local IGs and the global fit.

\begin{figure}
    \centering
    \includegraphics[width=0.85\columnwidth]{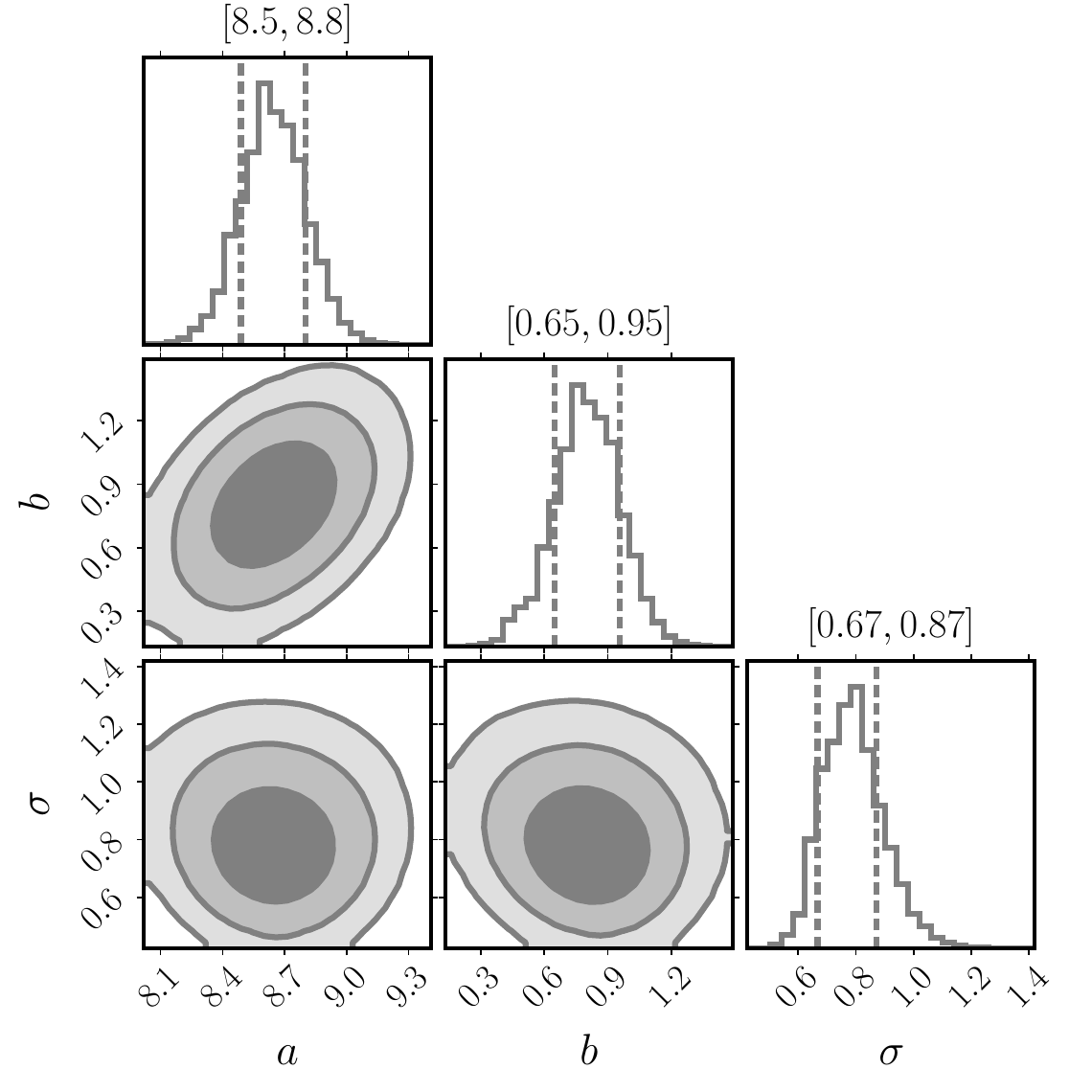}\\
    \includegraphics[width=0.85\columnwidth]{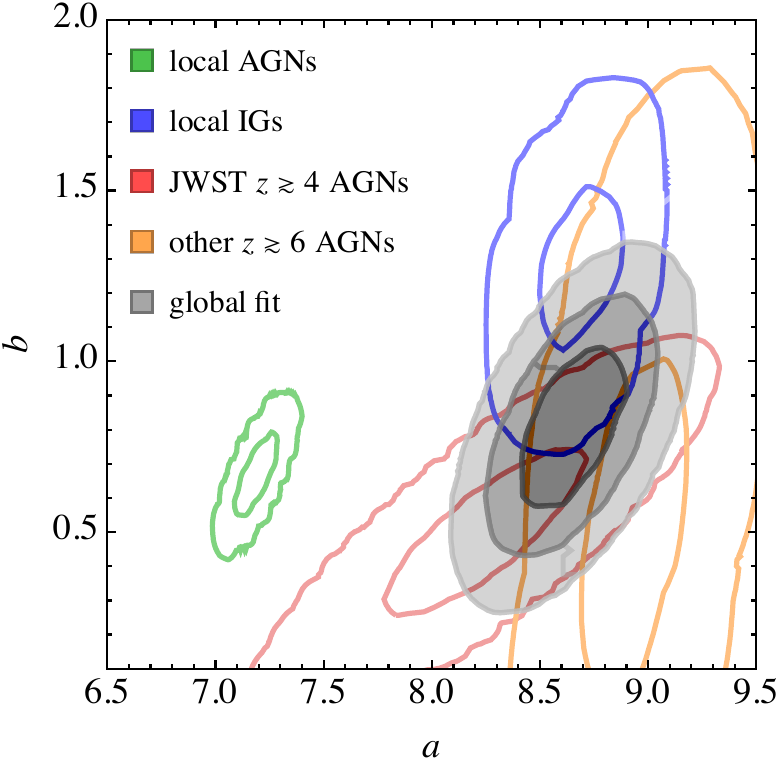}
    \caption{{\it Upper Panel}: Corner plot showing the posteriors for the global fit to JWST~\cite{2023A&A...677A.145U,2023ApJ...953L..29L,2023ApJ...959...39H,Bogdan:2023ilu,Ding_2023,Maiolino:2023bpi,2023arXiv230904614Y} and previous high-$z$ SMBH data~\cite{2021ApJ...914...36I} as well as low-$z$ data on SMBHs in IGs~\cite{2015ApJ...813...82R}. {\it Lower Panel}: The 68\%, 95\% and 99.8\% CL contours (gray) of the posteriors for $a$ and $b$ defined in~\eqref{eq:parameters}, obtained in a global fit including all observations except local AGNs by marginalizing over $\sigma$. For comparison, the 68\% and 99.8\% CL of individual data sets are shown: JWST high-$z$ measurements (red)~\cite{2023A&A...677A.145U,2023ApJ...953L..29L,2023ApJ...959...39H, Bogdan:2023ilu, Ding_2023, Maiolino:2023bpi, 2023arXiv230904614Y}, other high-$z$ measurements (yellow)~\cite{2021ApJ...914...36I}, local IGs (blue) and local AGNs (green)~\cite{2015ApJ...813...82R}. The fits are performed using flat priors and including the uncertainties in the observations. Corner plots showing the posterior distributions for these fits are displayed in the Appendix.}
    \label{fig:Ellipses}
\end{figure}

\begin{table}
\centering
\begin{tabular}{C{30mm}|C{11mm}C{11mm}C{10mm}C{10mm}C{10mm}}
    \hline\hline
    & JWST $z\gtrsim 4$ AGNs & other $z\gtrsim 6$ AGNs &local AGNs & local IGs & global fit \\ \hline
    JWST $z\gtrsim 4$ AGNs &  1 & \cellcolor{red!10}0.07 & \cellcolor{red!40}$10^{-6}$ & \cellcolor{red!15}0.007 & \cellcolor{red!5}0.3 \\
    other $z\gtrsim 6$ AGNs & & 1 & \cellcolor{red!50}$10^{-9}$&\cellcolor{red!10}0.02&\cellcolor{red!5}0.15 \\
    local AGNs & & & 1 &\cellcolor{red!50}$10^{-9}$ &\cellcolor{red!50}$10^{-9}$ \\
    local IGs & & & & 1&\cellcolor{red!10}0.08 \\
    global fit & & & & & 1\\
    \hline\hline
\end{tabular}
\caption{Overlaps~\eqref{eq:O} between the posteriors obtained in the fits to different data sets.
The cells of the Table are colour-coded for convenience.}
\label{tab:overlaps}
\end{table}

\begin{figure}
    \centering
    \includegraphics[width=\columnwidth]{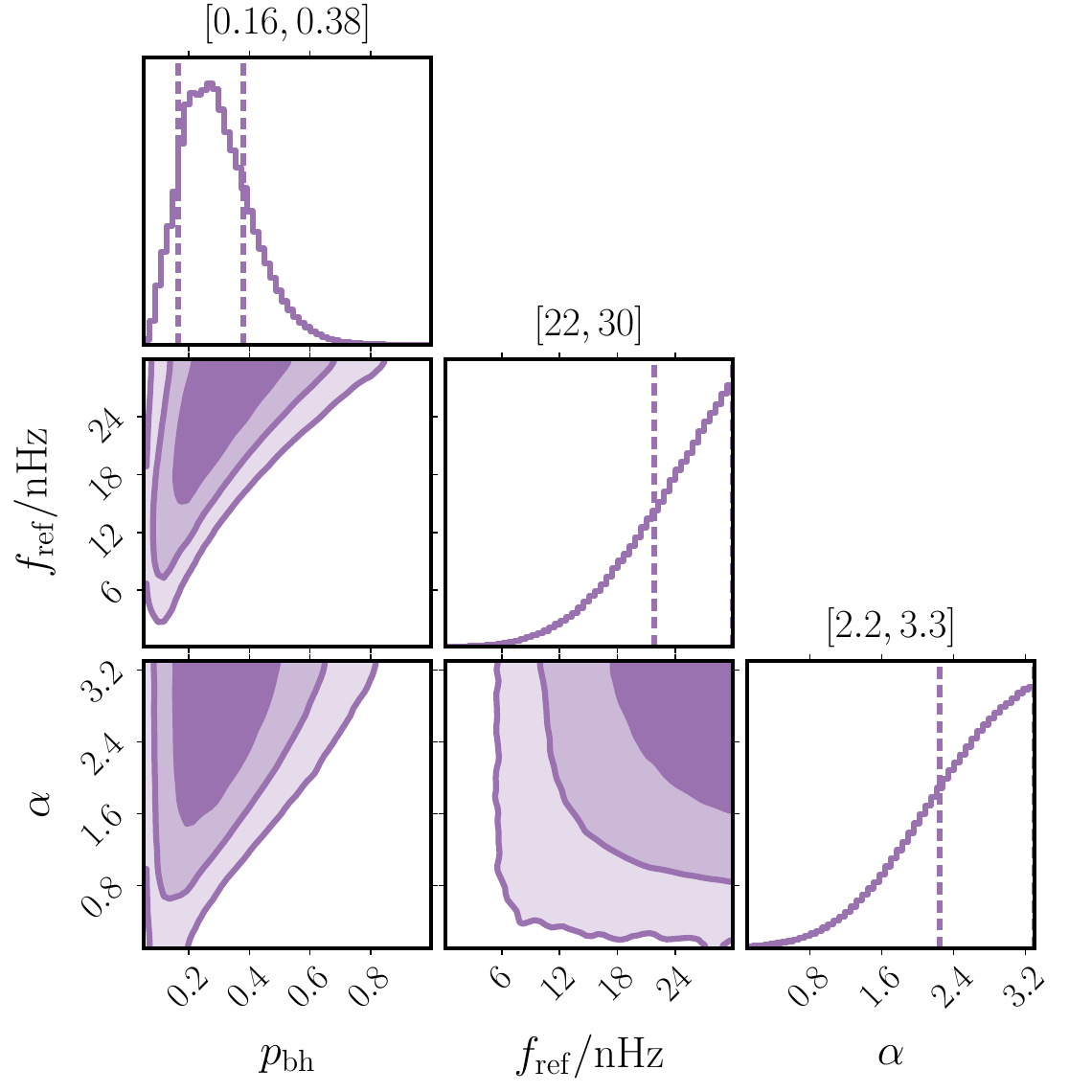}
    \caption{Posterior distributions of the parameters for a fit to NANOGrav GW data~\cite{NANOGrav:2023gor} including energy loss due to interactions with the environment.} 
    \label{fig:SMBH_post}
\end{figure}

In~\cite{Ellis:2023dgf} we found that the local IG fit also gives a much better fit to the NANOGrav GW observations than the local AGN fit, and we now develop this point further. As in~\cite{Ellis:2023owy, Ellis:2023dgf, Ellis:2023iyb}, we relate the SMBH merger rate to the halo merger rate $R_h$, estimated from the Press-Schechter formalism~\citep{Press:1973iz, Bond:1990iw, Lacey:1993iv}, as
\bea \label{eq:mergerrate}
    \frac{\td R_{\rm BH}}{\td m_1 \td m_2} \approx & \,\,p_{\rm BH} \int  \td M_{{\rm v},1} \td M_{{\rm v},2} \frac{\td R_h}{\td M_{{\rm v},1} \td M_{{\rm v},2}} \\
    &\hspace{10pt} \times \prod_{j=1,2} \frac{\td P(m_j|M_*(M_{{\rm v},j},z))}{\td m_j} \,, 
\eea
where $p_{\rm BH}$ characterizes the BH merger efficiency and $M_*(M_{{\rm v},j},z)$ is the halo mass-stellar mass relation that we take from~\cite{Girelli:2020goz}. We perform the analysis of the NANOGrav GW signal in the same way as in~\cite{Ellis:2023dgf}, including environmental effects parametrized by a spectral index $\alpha$ and a reference frequency $f_{\rm ref}$ (see Eq.~\eqref{eq:tint} and the surrounding discussion). Fig.~\ref{fig:SMBH_post} displays in a corner plot the posterior distributions for the parameters of this fit: $\alpha, f_{\rm ref}$ and $p_{\rm BH}$. The posterior for $\alpha$ is compatible with models of environmental effects and the fit prefers large values of $f_{\rm ref} \gtrsim 10$\,nHz. The preferred range of $p_{\rm BH}$ is $[0.16, 0.38]$. The best fit is achieved at $p_{\rm BH}=0.37$, $f_{\rm ref}=27\,$nHz and $\alpha=3.3$. In particular, the value $\alpha=8/3$ expected for gas infall agrees well with the data. By fixing $\alpha=8/3$, we find that the best fit is given by $p_{\rm BH} = 0.37$, $f_{\rm ref}=30\,$nHz.

\begin{figure}
    \centering
    \includegraphics[width=\columnwidth]{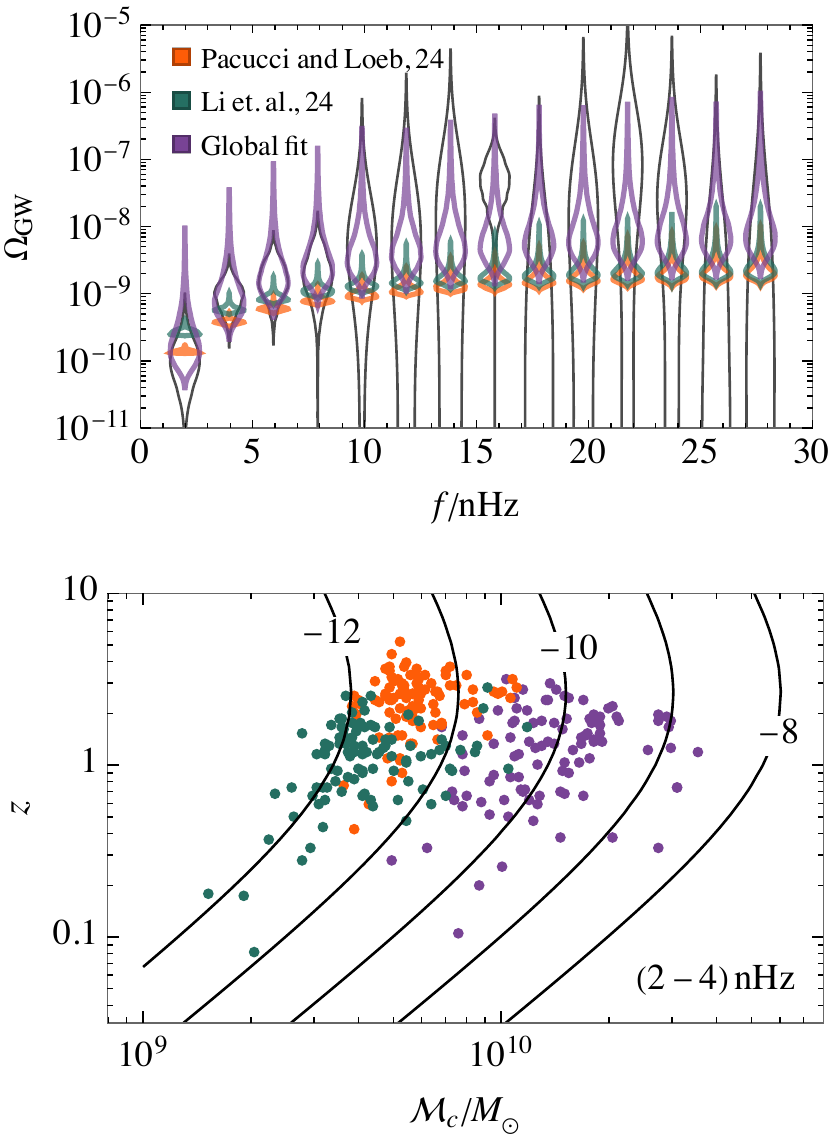}
    \caption{\textit{Upper panel:} Best fit nHz GW signals found using the BH mass-stellar mass relations proposed in~\cite{Pacucci:2024ijt}, \cite{2024arXiv240300074L} and our global fit~\eqref{eq:globalfit} compared to the signal measured by NANOGrav (gray). We imposed $\alpha=8/3$ as expected from gas infall (see discussion below Eq.~\eqref{eq:tint}). \textit{Lower panel:} Chirp masses and redshifts of 100 realizations of the strongest binary contributing to the first bin of the upper panel for the scenarios proposed in~\cite{Pacucci:2024ijt,2024arXiv240300074L}, compared with our analysis~\cite{Ellis:2023dgf}. Contours of $\log_{10}\Omega_{\rm GW}$ are shown for comparison.} 
    \label{fig:NANO}
\end{figure}

In the upper panel of Fig.~\ref{fig:NANO} we compare our global fit~\eqref{eq:globalfit} with the NANOGrav SGWB data~\cite{NANOGrav:2023gor} and the best fit SMBH binary GW signals obtained with the BH mass-stellar mass relations derived in~\cite{Pacucci:2023oci} and~\cite{2024arXiv240300074L}. The data and the best-fit simulations are represented by ``violins" whose widths represent the posteriors as functions of the GW density: following~\cite{Ellis:2023dgf}, we calculate the overlaps between the violins to assess the qualities of the fits. For both~\cite{Pacucci:2023oci} and \cite{2024arXiv240300074L} the best fits saturate at $p_{\rm BH}=1$, i.e., every galactic merger leads to a GW signal, which is an indication that the generated SGWB is insufficient. Comparing the best-fit likelihoods, we find that the PTA GW background favours the global fit mass relation at higher than the $3\sigma$ CL compared to the scaling relations proposed in~\cite{Pacucci:2023oci,2024arXiv240300074L}.

We show in the lower panel of Fig.~\ref{fig:NANO} 100 random realizations of the strongest binary contributing to the first NANOGrav bin, $f_{\rm GW} \in (2 - 4)$~nHz, for each of the three scenarios. In the case of~\cite{2024arXiv240300074L}, which advocated for a strong bias in the scaling relation, we can see that the SMBHs are too light, whereas in the case of~\cite{Pacucci:2024ijt}, which advocates for a steep $z$ dependence, the heavy SMBH are merging mainly at $z>1$, which also decreases the amplitude significantly.

Furthermore, in the lower panel of Fig.~\ref{fig:NANO} we also see that black holes with $m_{\rm BH}<10^9\, M_{\odot}$ are very unlikely to be among the strongest sources contributing to the SGWB. Given that the mass and $z$ range probed by NANOGrav coincides with the dynamically measured SMBH masses, the high merger efficiency necessary to explain the observed GW background implies that the local IG population is not significantly biased. The local AGNs lie in a lower mass range, and it is an open question whether that population is biased. However, given the consistency between the high-$z$ AGN measurements and the local IG measurements and the fact that at high $z$ the AGN fraction is close to unity but drops quickly at $z\lesssim 1$~\cite{Aird:2017cbs, Georgakakis:2017vvv}, it seems that the local AGN population reflects the intrinsic low-mass tail of the full BH mass-stellar mass relationship.

Comparing the local AGN population to the global fit in Fig.~\ref{fig:massrelations} we see that most of the local AGNs lie more than $2\sigma$ below the mean of the global fit. Hence, according to this analysis, the AGN fraction in the local SMBH population can only be $\lesssim 5\%$, in agreement with X-ray observations~\cite{Aird:2017cbs, Georgakakis:2017vvv}. To conclude, in contrast to~\cite{2024arXiv240300074L}, our analysis indicates that the local AGN population represents the tip of the iceberg of the local SMBH population. As discussed in~\cite{Ellis:2023dgf, Ellis:2023oxs}, our interpretation of the NANOGrav and JWST data would favour the observability of GW signals from BH binaries in space-borne laser interferometers (e.g., LISA~\cite{2017arXiv170200786A}, TianQin~\cite{Wang:2019ryf} and Taiji~\cite{Ruan:2018tsw})  and deciHz GW detectors (e.g., AION~\cite{Badurina:2019hst}, AEDGE~\cite{AEDGE:2019nxb} and DECIGO~\cite{Kawamura:2020pcg}).

\vspace{5pt}\noindent{\bf Dual AGN fraction --} We found in~\cite{Ellis:2023dgf,Ellis:2023oxs} that the quality of the SMBH fit to the NANOGrav data is improved by allowing for the evolution of the binary system to be affected by interaction with the environment as well as the emission of GWs. A similar conclusion was reached by the NANOGrav Collaboration~\cite{NANOGrav:2023hfp}. We discuss now the extent to which the unexpectedly high fraction of dual AGNs observed with JWST~\cite{2023arXiv231003067P} can be accommodated within our analysis of high-$z$ SMBH data and NANOGrav SGWB measurements including environmental effects. 

We estimate the numbers of dual AGNs and objects that we identify as single AGNs assuming that the collisions of galaxies trigger AGN activity. Environmental effects determine the evolution of the separations of the components in dual systems, so to estimate the numbers of dual AGNs and objects seen as single AGNs we start by characterizing the effective timescales of different environmental effects.

Several such effects are expected to contribute to the binary evolution in different ranges of the binary separation~\cite{Kelley:2016gse}, as illustrated in Fig.~\ref{fig:teff}. 
The effective timescale for the binary evolution is defined as $t_{\rm eff} \equiv |E|/\dot E$, where $E$ is the binary total energy. The effective timescale can be expressed as 
\be
    t_{\rm eff}^{-1} = t_{\rm GW}^{-1} + t_{\rm env}^{-1} \,,
\ee
where $t_{\rm GW}$ is the timescale of the GW driven binary evolution and $t_{\rm env}$ is the timescale of the environmental effects. For a circular orbit the binary total energy is given by $E = - m_1 m_2/d$ where $d$ denotes the separation of the BHs, so, in terms of the effective timescale, the separation evolves as 
\be
    \dot d = -d \,t_{\rm eff}^{-1} \,.
\ee
The dominant contribution to the binary evolution at small separations arises from GW emission, whose timescale is 
\be
    t_{\rm GW} = \frac{5d^4}{1024 \eta M^3} \approx \frac{14\,{\rm Myr}}{\eta} \left[ \frac{M}{10^9\Msun} \right]^{-3} \left[ \frac{d}{0.1\,{\rm pc}} \right]^4 \,,
\ee
where $\eta$ and $M$ denote the symmetric mass ratio and the total mass of the binary. However, other energy-loss mechanisms come into play at separations probed by PTAs and other observations.

At kpc separations, dynamical friction drives the binary evolution~\cite{Kelley:2016gse}. Its timescale is~\cite{2008gady.book.....B}
\be \label{eq:tdyn}
    t_{\rm dyn} \simeq \frac{20\,{\rm Myr}}{\ln \Lambda} \frac{\sigma}{200\,{\rm km}/{\rm s}} \left[\frac{M}{10^9\Msun}\right]^{-1} \left[\frac{d}{\rm kpc}\right]^2 \,,
\ee
where $\ln \Lambda$ is the Coulomb logarithm, $\sigma$ is the velocity dispersion of stars. 

Dynamical friction becomes inefficient at shorter distances when the binary binding energy becomes larger than the kinetic energy of the stars whose orbits pass close to the binary. The binary evolution is then driven by loss-cone scattering, whose timescale can be estimated as~\cite{Quinlan:1996vp}
\be \label{eq:tsh}
    t_{\rm sh} \simeq \frac{\sigma}{H \rho d} \approx 300\,{\rm Myr} \frac{\sigma}{200\,{\rm km}/{\rm s}} \!\left[\frac{\rho}{10\Msun/{\rm pc}^3}\right]^{-1} \!\left[\frac{d}{\rm pc}\right]^{-1} \!\!,
\ee
where $H\approx 15-20$ is the dimensionless hardening rate and $\rho$ and $\sigma$ denote the mass density and the velocity dispersion of stars in the neighbourhood of the binary. 

At intermediate scales between the GW-driven and loss-cone scattering-driven phases, as mentioned earlier we follow~\cite{Ellis:2023dgf} by parametrizing the environmental effects with a power-law 
\be \label{eq:tint}
    t_{\rm int} = t_{\rm GW} \left[\frac{d}{d_{\rm ref}}\right]^{-\frac{3\alpha}{2}} .
\ee
The reference separation $d_{\rm ref}$ is related to the reference frequency used in~\cite{Ellis:2023dgf} by $d_{\rm ref}^3 = 8 M/(\pi f_{\rm ref} (\mathcal{M}/10^9\Msun)^{-\beta})^2$. We estimate the total timescale of the environmental effects as $t_{\rm env} = (t_{\rm int}^{-1} + t_{\rm sh}^{-1})^{-1} + t_{\rm dyn}$. 

Effects that potentially determine the dynamics of binaries at intermediate scales include gas infall and viscous drag. The timescale of the binary evolution by gas infall can be estimated from the accretion timescale~\cite{1980Natur.287..307B}, $t_{\rm gas} \simeq m_1/\dot m_1$, where $m_1$ is the mass of the heavier BH, and the timescale of viscous drag has been estimated in~\cite{1999MNRAS.307...79I} to scale as, $t_{\rm vd} \propto \sqrt{d}/M^{1/14}$. These correspond, respectively, to $(\alpha,\beta) = (8/3,5/8)$ and $(\alpha,\beta) \approx (7/3,3/4)$, while a transition directly from the loss-cone scattering to GW-driven evolution would correspond to $(\alpha,\beta) = (10/3,2/5)$. For the PTA fit shown in Fig.~\ref{fig:SMBH_post} we fixed $\beta = 5/8$ after checking that the fit is not very sensitive to the value of $\beta$. We also find that the above values of $\alpha$ provide good fits to the data. 

Given the estimates for the effective timescales, we can estimate the expected number of dual and single AGNs detected by JWST. We assume that AGN activity is triggered only during major mergers where the stellar mass of the heavier galaxy is not more than three times the stellar mass of the lighter one. In minor mergers only the SMBH in the lighter galaxy typically becomes active and its activity ceases quickly as the larger galaxy rips apart the smaller one. This is in agreement with numerical simulations~\cite{Scudder:2012ut,Callegari:2010gc,Capelo:2014gqa} and low-$z$ observations~\cite{Treister:2012ag,Comerford:2015qda,Donley_2018,2012ApJ...746L..22K}, except for a small number of objects~\cite{Liu_2018,Micic:2023oaf}. Dual AGNs in which one of the AGNs is not luminous enough or whose angular separation is not large enough to be resolved and AGNs generated in mergers where the other galaxy did not contain an SMBH are seen as single AGNs.

\begin{figure}
    \centering
    \includegraphics[width=\columnwidth]{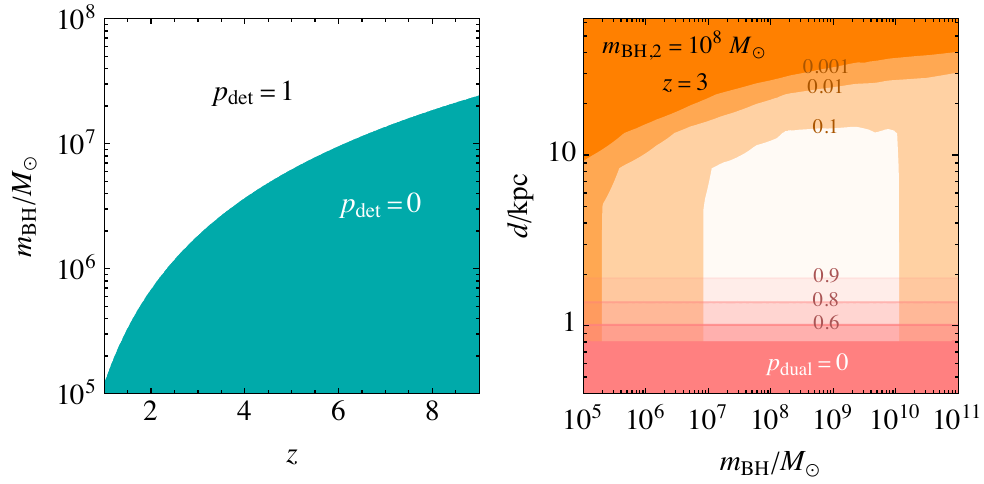}
    \caption{\emph{Left panel:} The detectability probability $p_{\rm det}$ arising from the JWST sensitivity threshold on the flux. \emph{Right panel:} The red regions show the dual identification probability $p_{\rm dual}$ at $z=3$ arising from the JWST angular resolution. The orange regions show the probability that both galactic nuclei are active for $m_{\rm BH,2} = 10^8\Msun$ at $z=3$.}
    \label{fig:pdetpdual}
\end{figure}

\begin{figure}
    \centering
    \includegraphics[width=\columnwidth]{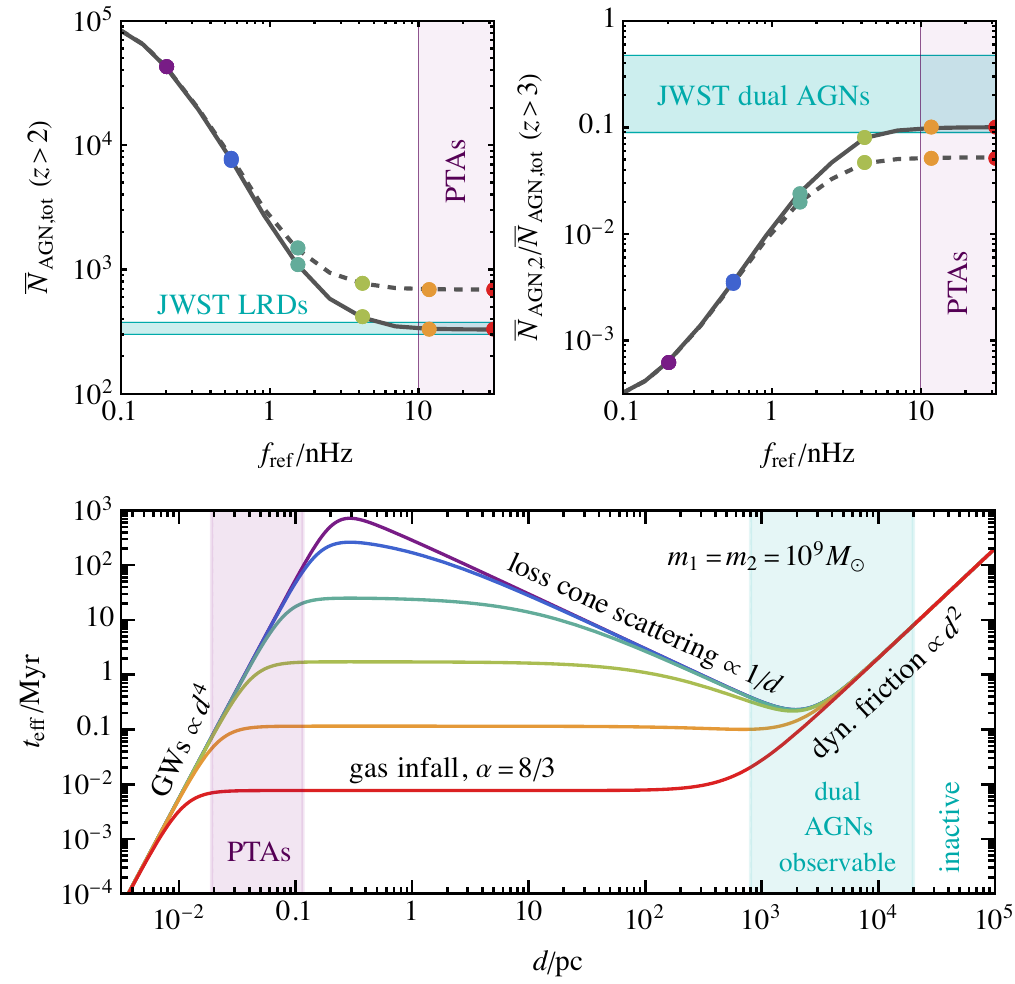}
    \caption{\textit{Upper panels:} The expected total number of AGNs at $z>2$ and the expected dual fraction at $z>3$ as a function of the reference frequency below which the binaries are driven by GW emission, assuming $p_{\rm BH} = 0.2$, $\alpha=8/3$, $A_{\rm dyn} = 0.001$ and $A_{\rm sh} = 1$. The dashed and solid curves are with and without galaxy mergers where one of the galaxies does not contain an SMBH. \textit{Lower panel:} Illustration of the timescales of the environmental energy loss mechanisms of SMBH binaries as functions of their separations for $\alpha=8/3$, $A_{\rm dyn} = 10^{-3}$ and $A_{\rm sh} = 1$. The vertical bands illustrate the separation ranges probed by the PTAs and by the JWST dual AGN observations. The colour coding of the curves matches that shown in the upper panels.}
    \label{fig:teff}
\end{figure}

The expected number of detectable dual AGNs is
\be
    \bar{N}_{\rm AGN,2} = f_{\rm sky} \int \!\td \lambda_{\rm 2}\, p_{\rm det}(m_1,z) p_{\rm det}(m_2,z) p_{\rm dual}(d,z)  \,,
\ee
where $f_{\rm sky}$ is the sky fraction being observed, $p_{\rm det}$ the probability that the BH is luminous enough to be detectable, and $p_{\rm dual}$ is the probability that a dual AGN is resolvable. Counting all mergers where at least one of the galaxies is occupied by an SMBH that becomes active during the merger and is sufficiently luminous to be detected, the expected number of detectable AGNs is
\bea \label{eq:N_1}
    \bar{N}_{\rm AGN,tot} =& \,\, 2 f_{\rm sky} \int \td \lambda_1 \, p_{\rm det}(m,z) \\ 
    &+ 2 f_{\rm sky} \int \td \lambda_2 \, p_{\rm det}(m_2,z) \\
    &- f_{\rm sky} \int \td \lambda_{2} \, p_{\rm det}(m_1,z) p_{\rm det}(m_2,z) \,.
\eea
The merger rate and the evolution of the BH separation enter through
\be
    \td \lambda_1 
    \equiv \frac{1}{1+z} \frac{\td V_c}{\td z} \frac{\td n_{\rm AGN}^{1}}{\td m \td d} \td m \td z \td d
\ee
and 
\be
    \td \lambda_2 
    \equiv \frac{1}{1+z} \frac{\td V_c}{\td z} \frac{\td n_{\rm AGN}^{2}}{\td m_1 \td m_2 \td d} \td m_1 \td m_2 \td z \td d \,,
\ee
where 
\bea \label{eq:nAGN1}
    &\frac{\td n_{\rm AGN}^1}{\td m \td d}
    = \frac{1-\sqrt{p_{\rm BH}}}{|\dot d|} \int \!\td M_{{\rm v},1} \td M_{{\rm v},2} \frac{\td R_h}{\td M_{{\rm v},1} \td M_{{\rm v},2}} \\ 
    &\times p_{\rm AGN}(M_{{\rm v},1},M_{{\rm v},2},d) \frac{\td p_{\rm occ}(m|M_*(M_{{\rm v},2},z))}{\td m}
\eea
is the number density of AGNs formed in galaxy mergers where one of the galaxies did not contain an SMBH, and
\bea \label{eq:nAGN2}
    &\frac{\td n_{\rm AGN}^{2}}{\td m_1 \td m_2 \td d}
    = \frac{1}{|\dot d|} \int \!\td M_{{\rm v},1} \td M_{{\rm v},2} \frac{\td R_h}{\td M_{{\rm v},1} \td M_{{\rm v},2}} \\ 
    &\!\times\! p_{\rm AGN}(M_{{\rm v},1},M_{{\rm v},2},d) \!\!\prod_{j=1,2} \! \frac{\td p_{\rm occ}(m_j|M_*(M_{{\rm v},j},z))}{\td m_j}
\eea
is the number density of dual AGNs. The probability that a galaxy with stellar mass $M_*$ is occupied by an SMBH with mass $m$ can be expressed as 
\be
    \frac{\td p_{\rm occ}(m|M_*)}{\td m} = \sqrt{p_{\rm BH}} \,\frac{\td P(m|M_*)}{\td m}
\ee
where the distribution $\td P/\td m$, given by Eq.~\eqref{eq:parameters}, is normalized to 1, so $\sqrt{p_{\rm BH}}$ can be interpreted as the SMBH occupation fraction, which we take to be independent of $M_*$. The prefactor $1-\sqrt{p_{\rm BH}}$ in~\eqref{eq:nAGN1} gives the probability that one of the merging galaxies does not contain an SMBH. In this case, dynamical friction drives this single SMBH towards the centre of the galaxy that resulted from the merger. So, for Eq.~\eqref{eq:nAGN1} we estimate the timescale as $t_{\rm eff} = t_{\rm dyn}$. We assume that the AGN activity starts when the distance between the galaxies becomes smaller than the half-stellar radius of the smaller galaxy. We approximate the half-stellar radius by $10\%$ of the virial radius of the halo:
\bea
    &p_{\rm AGN}(M_{{\rm v},1},M_{{\rm v},2},d) 
    = \Theta(0.1 \min_j R_{{\rm v},j} - d) \\
    &\hspace{40pt} \times \Theta(3 \min_j M_{*,j} - \max_j M_{*,j}) \,.
\eea
In the right panel of Fig.~\ref{fig:pdetpdual} the shades of orange show the dual AGN probability $\td n_{\rm AGN}^2/\td n_{\rm tot}$ where $\td n_{\rm tot}$ is obtained from Eq.~\eqref{eq:nAGN2} with $p_{\rm AGN} = 1$.

We estimate the detection probability $p_{\rm det}$ by setting a threshold on the flux
\be
    p_{\rm det}(m_{\rm BH},z) = \Theta\left(\frac{L}{4\pi D_L^2}-F_{\rm sens}\right) \,,
\ee
where $D_L$ denotes the luminosity distance of the system. We approximate the AGN luminosity optimistically as $L \simeq L_{\rm Edd}/3 \approx 4.3\times 10^{43} \,(m_{\rm BH}/10^6\Msun) \,{\rm erg}/{\rm s}$ and choose $F_{\rm sens} \simeq 10^{-15} \,{\rm erg}/{\rm s}/{\rm cm}^2$, which corresponds to the dimmest source found in~\cite{2023arXiv231003067P}. The threshold on the flux gives a redshift-dependent lower bound on the BH mass, as shown in the left panel of Fig.~\ref{fig:pdetpdual}.

We estimate the probability $p_{\rm dual}$ by assuming that the duality of the AGNs is properly identified if its projected angular separation,
\be
    \theta(d,z) = \frac{d}{D_A(z)} \sqrt{1-\sin^2 i \sin^2\phi} \,,
\ee
where $D_A(z)$ denotes the angular diameter distance of the system and $i$ and $\phi$ the inclination and phase of the dual, is larger than twice the angular resolution $\theta_{\rm res}\approx 0.1''$ of JWST, and we average over the angles $i$ and $\phi$:
\be
    p_{\rm dual}(d,z) = \frac{1}{4\pi} \int \td i \,\td \phi \sin i \,\Theta(\theta(d,z)-2\theta_{\rm res}) \,.
\ee
The probability $p_{\rm dual}$ is shown in the right panel of Fig.~\ref{fig:pdetpdual}. 

We write the timescales of dynamical friction and stellar loss-cone scattering as $t_{\rm dyn} = 20\,{\rm Myr}\, A_{\rm dyn} [M/10^9\Msun]^{-1} [d/{\rm kpc}]^2$ and $t_{\rm sh} = 300\,{\rm Myr}\, A_{\rm sh} [d/{\rm pc}]^{-1}$ (see Eqs.~\eqref{eq:tdyn} and \eqref{eq:tsh}), and study how the dual AGN fraction $\bar{N}_{\rm AGN,2}/\bar{N}_{\rm AGN,tot}$ and the total number of AGNs $\bar{N}_{\rm AGN,tot}$ depend on the parameters $A_{\rm dyn}$, $A_{\rm sh}$, $p_{\rm BH}$, $f_{\rm ref}$ and $\alpha$. We consider the observations reported in~\cite{2023arXiv231003067P} of the GA-NIFS sample of 17 known AGNs at $z>3$ of which $N_1 = 12$ were identified as single AGNs, $N_2 = 4$ as dual AGNs and one as a triple AGN. Based on these observations, we estimate that the dual fraction $N_2/(N_1+N_2) = 0.25$.  We also consider the JWST results of~\cite{2024arXiv240403576K} that cover $587.8 \,{\rm arcmin}^2$, corresponding to $f_{\rm sky} \approx 4\times 10^{-6}$, revealing 341 ``little red dots" (LRDs) in the redshift range $2 < z < 11$. Interpreting these LRDs as AGNs, we use this as an observation of the total AGN number $N_{\rm tot} = 341$.\footnote{The origin of the LRDs is debated because they are not detected in X-rays~\cite{Ananna:2024jug}. However, the non-detection of X-rays may be caused by high obscuration or intrinsic X-ray weakness of the AGNs~\cite{Maiolino:2024uon}.}

In the upper panels of Fig.~\ref{fig:teff} we show the dependencies of the numbers of AGNs and fractions of dual AGNs on $f_{\rm ref}$ for $\alpha=8/3$, corresponding to the gas infall timescale, $A_{\rm sh} = 1$, $A_{\rm dyn} = 0.001$ and $p_{\rm BH} = 0.2$. For these parameters, the lower panel of Fig.~\ref{fig:teff} illustrates the timescales of the environmental energy loss mechanisms, including GW emission, gas effects, stellar loss-cone scattering and dynamical friction, as functions of the binary separation. The different curves in the lower panel correspond to different values of $f_{\rm ref}$ indicated by the coloured dots in the upper panel. The horizontal bands in the upper panel show the 95\% CL bands of the JWST LRD and dual AGN observations calculated assuming the Poisson and binomial distribution, respectively. The value of $p_{\rm BH}$ is fixed so that, neglecting AGNs that may form in mergers where only one of the galaxies contains an SMBH, the expected total AGN number roughly matches the LRD observations for large $f_{\rm ref}$. We see in the right panel that, in this benchmark case, the dual fraction is within the 95\% range of the observed value for $f_{\rm ref} > 6$\, nHz if only the AGNs that may form in mergers where both of the galaxies contain an SMBH are considered. These values of $p_{\rm BH}$, $\alpha$ and $f_{\rm ref}$ are within the 95\% CL region of the NANOGrav fit, illustrating that it is possible to obtain a coherent description of the evolution of binary systems that explains both the NANOGrav GW signal and the JWST dual AGN and LRD observations.

Fig.~\ref{fig:teff} shows that the expected number of systems seen as single AGNs would be much larger than observed if $f_{\rm ref}$ is small, because in that case, the binaries would spend a lot of time orbiting with $d\sim 1$\,pc separation. This is related to the well-known final-parsec problem. Both the JWST and PTA observations prefer large values of $f_{\rm ref}$ for which the final-parsec problem is avoided. Then, the main contribution to the total number of AGNs comes from the same region as for dual AGNs, $d\gtrsim 1$\,kpc, where the effective timescale is largest. In this regime, both $\bar{N}_{\rm AGN,2}$ and $\bar{N}_{\rm AGN,tot}$ are proportional to the prefactor $A_{\rm dyn}$ of the dynamical friction timescale. For the benchmark shown in Fig.~\ref{fig:teff} we have fixed $A_{\rm dyn} = 0.001$ so that the result is consistent with the LRD observations for values of $p_{\rm BH}$ consistent with the NANOGrav observations.

\vspace{5pt}\noindent{\bf Conclusions --}  We have presented a global analysis of the available data on high-$z$ SMBHs detected with the JWST~\cite{2023A&A...677A.145U,2023ApJ...953L..29L,2023ApJ...959...39H,Bogdan:2023ilu,Ding_2023,Maiolino:2023bpi,2023arXiv230904614Y} and in previous observations~\cite{2021ApJ...914...36I} in conjunction with results from previous samples of SMBHs in low-$z$ AGNs and IGs~\cite{2015ApJ...813...82R} and measurements of the SGWB background by NANOGrav~\cite{NANOGrav:2023gor}, which are compatible with data from other PTAs~\cite{InternationalPulsarTimingArray:2023mzf}. The JWST data are compatible with previous high-$z$ SMBH data, and both are compatible with a sample of low-$z$ IGs, but are quite incompatible with a low-$z$ AGN sample -- see Fig.~\ref{fig:Ellipses} and Table~\ref{tab:overlaps}. Moreover, we have shown that our global fit to the high-$z$ SMBH and local IG data also fits well the NANOGrav data on the nHz SGWB, whereas alternative interpretations of the high-$z$ SMBH data that invoke evolution with $z$ of the $M_{\rm BH}/M_{\rm *}$ relationship~\cite{Pacucci:2024ijt} or selection biases and measurement uncertainties~\cite{2024arXiv240300074L} provide only poor fits to the SGWB data, as seen in Fig.~\ref{fig:NANO}.

We have also shown how the dual AGN fraction reported in~\cite{2023arXiv231003067P} can be accommodated within our global analysis. Our interpretation of the data on the dual AGN fraction~\cite{2023arXiv231003067P} is related to the need for environmental effects to accommodate the NANOGrav data, see Fig.~\ref{fig:SMBH_post} and the upper panel of Fig.~\ref{fig:NANO} as well as~\cite{NANOGrav:2023hfp} and~\cite{Ellis:2023dgf}. Fig.~\ref{fig:teff} shows that an extrapolation of these environmental effects to larger scales can accommodate the fraction of dual AGNs detected with JWST. Moreover, we also find that our global analysis is compatible with the abundance of ``little red dots" in JWST observations reported recently~\cite{Matthee:2023utn,2024arXiv240403576K}.

Observations of high-$z$ SMBHs are still in their infancy, with much more data to come from JWST and other sources. Likewise, PTA observations of the nHz SGWB will develop rapidly in the near future. Doubtless, both sets of data will provide new puzzles. The central point of our paper is that these data sets should be considered together and that these puzzles may share common features. Resolution of these puzzles may also require inputs from other observational programmes such as measurements of GWs at higher frequencies, interpreted within an integrated approach.

\begin{acknowledgments}
\vspace{5pt}\noindent\emph{Acknowledgments --}
This work was supported by the European Regional Development Fund through the CoE program grant TK202 and by the Estonian Research Council grants PRG803, PSG869, RVTT3 and RVTT7. The work of J.E. and M.F. was supported by the United Kingdom STFC Grants ST/T000759/1 and ST/T00679X/1. The work of V.V. was partially supported by the European Union's Horizon Europe research and innovation program under the Marie Sk\l{}odowska-Curie grant agreement No. 101065736.
\end{acknowledgments}

\bibliography{refs}

\newpage
\appendix
\onecolumngrid
\newpage
\section{Appendix: Compilation of JWST Data and Corner Plots}
\vspace{-5mm}
\begin{figure*}[h]
    \centering
    \includegraphics[width=0.41\textwidth]{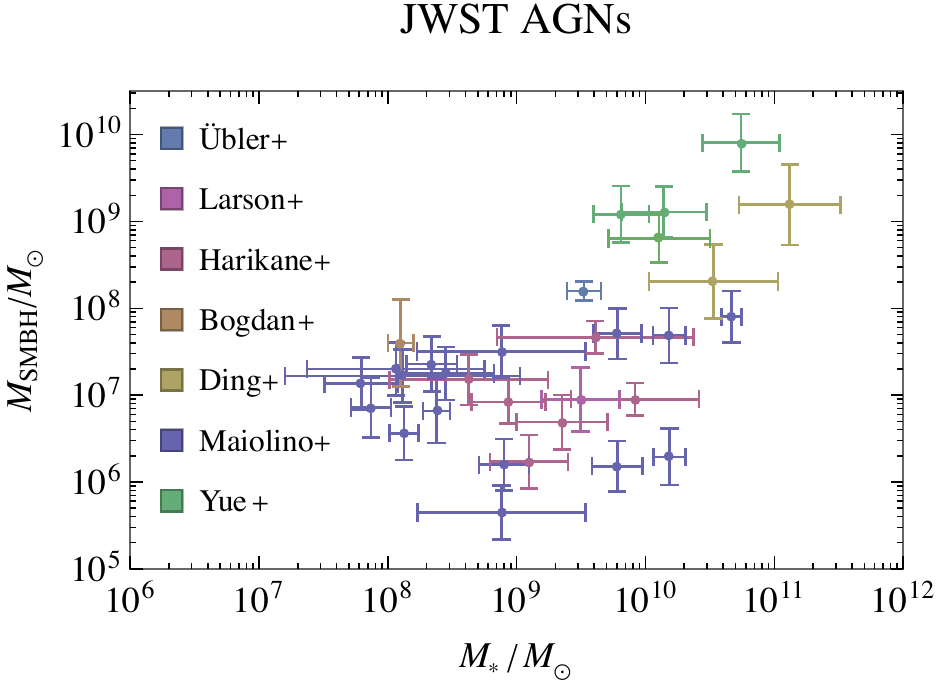}
    \caption{Scatter plot in the $(m_{\rm BH}/M_\odot, M_*/M_\odot)$ plane of the JWST data analyzed in this paper.
    The following are the papers listed in the legend: {\"U}bler+~\cite{2023A&A...677A.145U}, Larson+~\cite{2023ApJ...953L..29L}, Harikane+~\cite{2023ApJ...959...39H}, Bogdan+~\cite{Bogdan:2023ilu}, Ding+~\cite{Ding_2023}, Maiolino+~\cite{Maiolino:2023bpi}, Yue+~\cite{2023arXiv230904614Y}.}
    \label{fig:extra_info}
\end{figure*}
\vspace{-5mm}

%\section{Appendix B: Corner plots}
~~\\
We present below corner plots that exhibit the posteriors for the parameters $a$, $b$ and $\sigma$ that were introduced in~\eqref{eq:parameters} to characterize the BH mass-stellar mass relation for fits to various data sets discussed in the main text. These include (upper left) the JWST high-$z$ SMBH data compiled in Fig.~\ref{fig:extra_info}, (upper right) the previous high-$z$ SMBH data compiled in~\cite{2021ApJ...914...36I}, (lower left) the combined high-$z$ SMBH data,  and (lower right) the combination of these data with the RV data on SMBHs in inactive galaxies (IGs)~\cite{2015ApJ...813...82R}.

\begin{figure*}[h]
    \centering
    \includegraphics[width=0.31\textwidth]{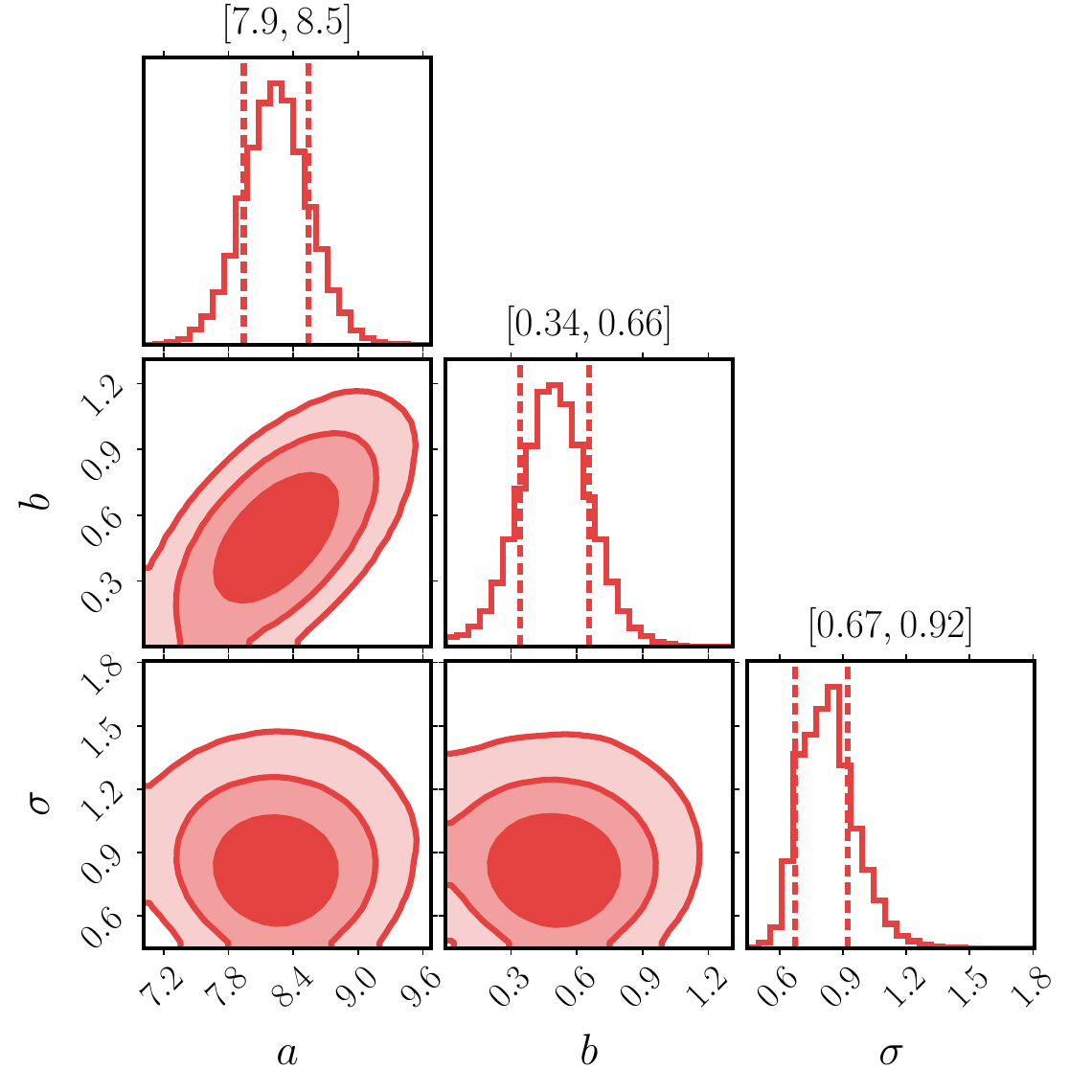}
    \includegraphics[width=0.31\textwidth]{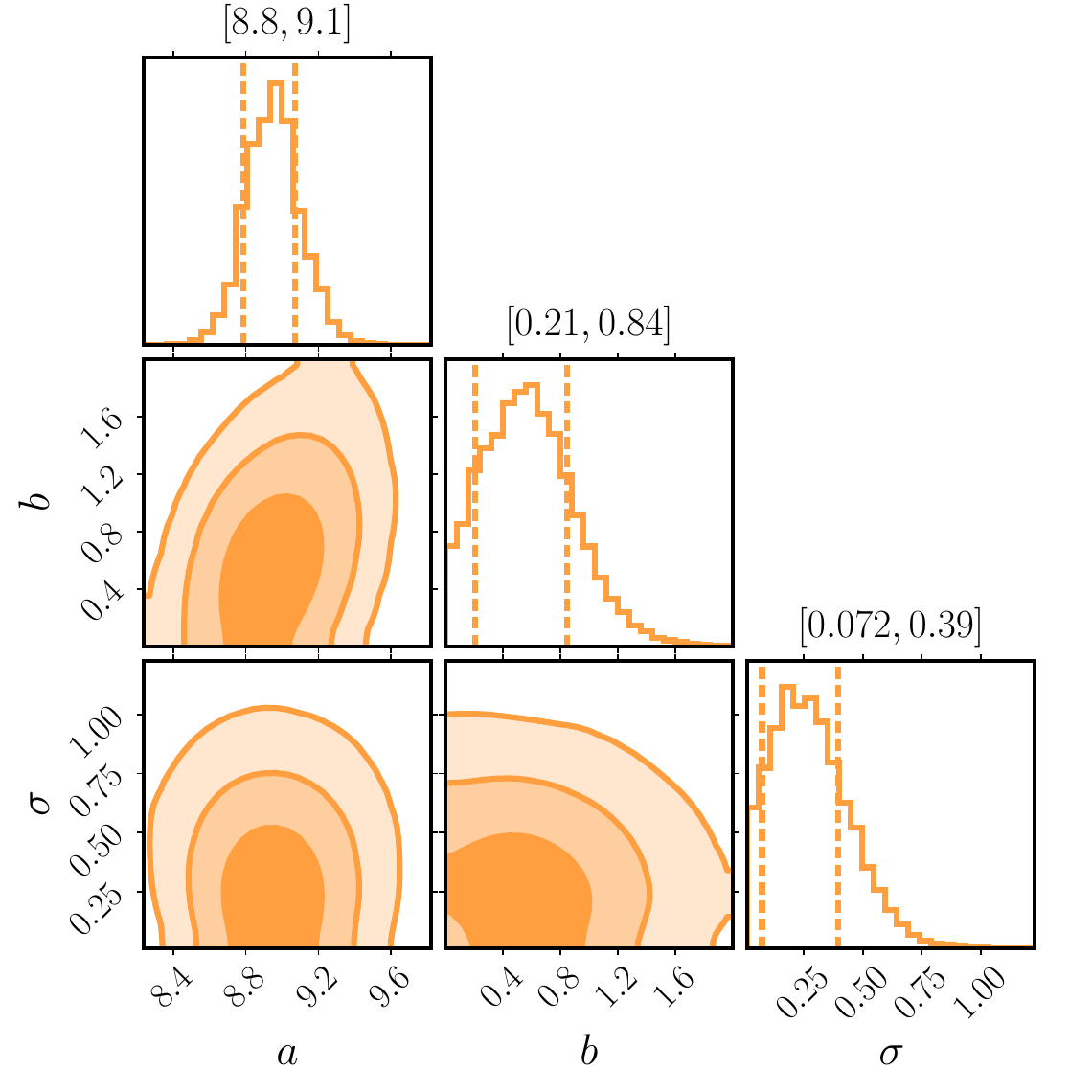} \\
    \includegraphics[width=0.31\textwidth]{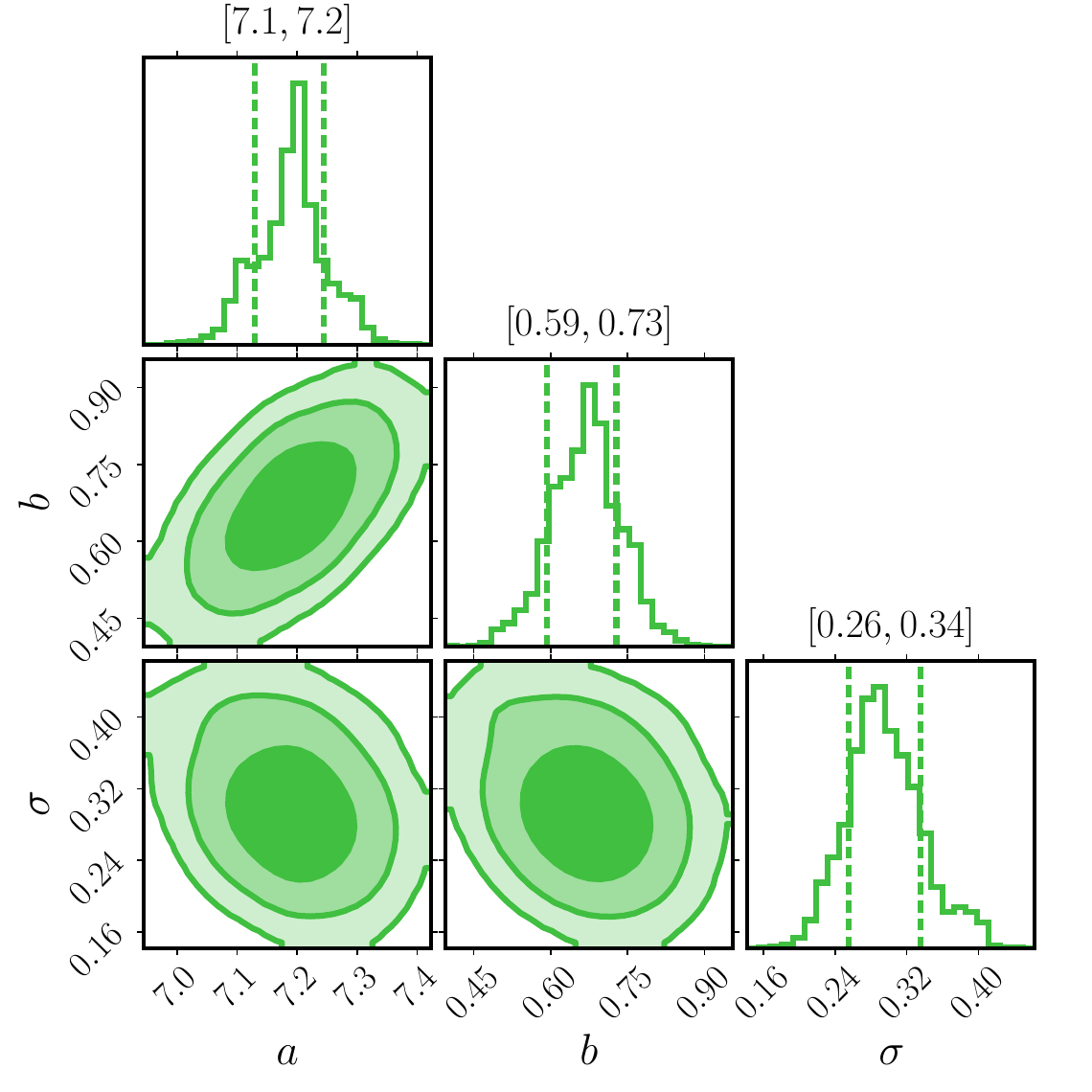}
    \includegraphics[width=0.31\textwidth]{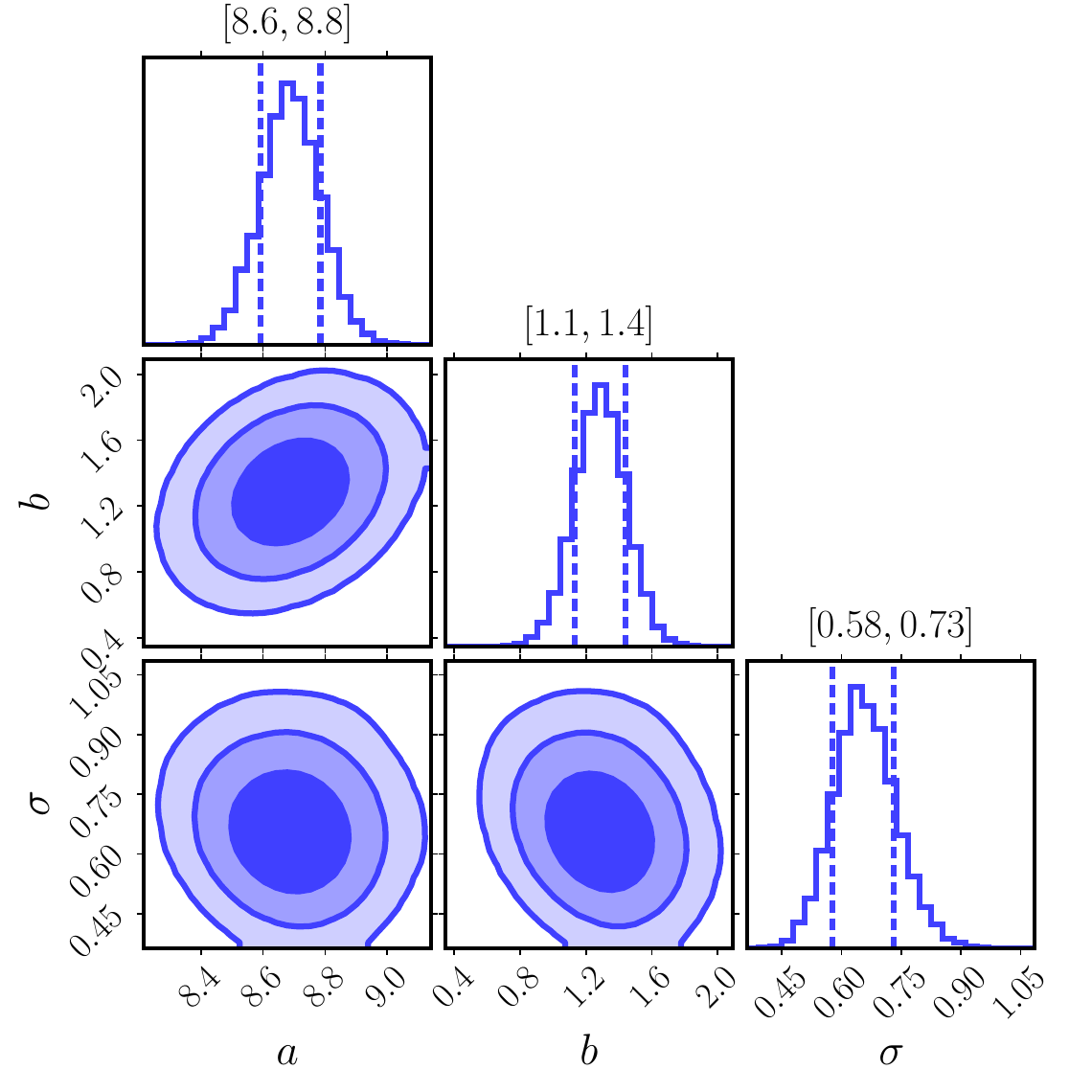}
    \caption{Posteriors for fits to \textit{(upper left)} the JWST high-$z$ SMBH data compiled in Fig.~\ref{fig:extra_info}, \textit{(upper right)} previous high-$z$ SMBH data compiled in~\cite{2021ApJ...914...36I}, \textit{(lower left)} the RV data on local AGNs~\cite{2015ApJ...813...82R}, and \textit{(lower right)} the dynamically measured SMBH compiled in~\cite{2015ApJ...813...82R}.}
    \label{fig:Fits+Subaru}
\end{figure*}
\end{document}